\documentclass[aps,prl,twocolumn,superscriptaddress,nofootinbib,longbibliography]{revtex4-2}
\usepackage[utf8]{inputenc}
\usepackage[hidelinks,breaklinks]{hyperref}

\usepackage{graphicx}
\usepackage[shortlabels]{enumitem}

\usepackage{braket}
\usepackage{amssymb}
\usepackage{mathtools}

\usepackage{xcolor}

\begin{document}

\title{Distinguishing Bohmian contextuality from Kochen-Specker contextuality}

\author{Anton Skott}
\email{antonsk@stanford.edu}
\affiliation{Department of Philosophy, Stanford University, Stanford CA, USA}
\author{Jan-{\AA}ke Larsson}
\email{jan-ake.larsson@liu.se}
\affiliation{Department of Electrical Engineering, Link\"oping University, 581 83 Link\"oping, SWEDEN}

\begin{abstract}
\sloppy
Quantum contextuality is a concept used to describe the property of hidden-variable theory that measurement outcomes predetermined by the hidden variables depend on the measurement context.
The term measurement context can have different meanings, giving rise to different flavours of quantum contextuality.
The first discovered flavour is Kochen-Specker (KS) contextuality where measurement outcomes will depend on what compatible measurements are jointly performed with the selected measurement.
Another flavour, here to be compared with KS contextuality, is that referred to in Bohmian mechanics where outcomes of some specific measurements are not completely specified by the model state, but depend also on specifics of the measurement device used.
It has been claimed that this type of Bohmian contextuality is necessary to enable KS contextuality in a hidden variable model.
In this paper we show that this is not the case.
The recently proposed Contextual Ontological Model (COM) [Hindlycke\&Larsson, Phys.~Rev.~Lett.~2022] produces KS contextual predictions but does not have the Bohmian contextuality; the outcome of every measurement allowed by COM can be predicted from the model state itself.
This distinguishes Bohmian contextuality from KS contextuality, and enables individual study of the two concepts.
\end{abstract}
\maketitle

\section{Introduction}
A central concept in discussions of the foundations of quantum mechanics is contextuality. On the physics side of the literature, contextuality is usually understood in the formal sense given by the Kochen-Specker (KS) theorem \cite{Kochen1967,Budroni2022} which places constraints on hidden-variable theories.
If a hidden-variable theory is to reproduce quantum-mechanical predictions, then it must necessarily be KS contextual, meaning that the prediction must depend not only on the performed measurement but also its context, the compatible measurements that are jointly performed with the target measurement.
In other parts of the literature and in particular that concerning Bohmian mechanics, contextuality is sometimes understood as the dependence of measurement outcomes on details of the measurement setup itself \cite{Albert1992,Hardy1996,Norsen2014}. 
These two notions of contextuality are similar but not identical. In this paper, we seek to clarify this relation.

A hidden-variable theory that exhibits both types of contextuality is Bohmian mechanics \cite{Bohm1952,Holland1993}. 
As is previously known and will be clarified here using sequential Stern-Gerlach  measurements, the outcome of a measurement in Bohmian mechanics depends both on the initial state of the quantum system and the measurement setup. 
Implicitly or explicitly, this form of dependency is sometimes argued to be required for Kochen-Specker contextuality in quantum mechanics generally (see e.g.,~\cite{Norsen2014}). 

In this paper, we compare Bohmian mechanics with the Contextual Ontological Model \cite{Hindlycke2022,Larsson2026} (COM).
COM is a hidden-variable model written in the language of stabilizer quantum mechanics \cite{Gottesman1998b,Aaronson2004a} that exhibits KS contextuality.
However, in COM the measurement outcomes are predicted directly from the model state, and no details of the measurement are needed; the preexisting measurement outcomes are simply revealed in a measurement. 
Therefore Bohmian contextuality as discussed in Refs.~\cite{Albert1992,Hardy1996,Norsen2014} is not present, meaning that COM is a relatively simple model that has KS contextuality, even though the Bohmian form of contextuality is absent. 
This directly implies that Bohmian contextuality is not needed for KS contextuality, which clarifies the relation between these two different notions of contextuality; they are separate notions. 
It also stresses the importance of distinguishing between notions of contextuality that could easily be and often are conflated in both the physics and philosophy literature. 

The analysis uses the framework of COM which is restricted to predictions from stabilizer quantum mechanics. 
But in the present analysis this restriction should rather be seen as a simplification that enables us to clarify the separation between the different notions of contextuality. 
That such a separation is possible even in our simplified setting shows that full quantum mechanics also does not force a connection between these notions, but that the notions are separate and should be analyzed separately.

A final interesting point is that our result shows that COM has different properties than Bohmian mechanics. 
COM does not exhibit Bohmian contextuality, while Bohmian mechanics does.
This implies that COM is not just a simplified version of Bohmian mechanics.
Rather, it is an independent hidden-variable model that could be developed further as an alternative ontological version of quantum mechanics. 

In what follows we will first introduce Bohmian mechanics and then delineate the spin-$\tfrac12$ measurement most often used to argue for measurement dependence as a basis for contextuality in Bohmian mechanics.
After this we will give a brief introduction to stabilizer quantum mechanics, and COM, and finally study a spin-$\tfrac12$ measurement within COM, and why it does not need this kind of measurement dependence.

\section{Bohmian Mechanics}
\label{cha:bohmian}

We will here follow the presentation of Bohmian mechanics found in Refs.~\cite{Norsen2014,Naaman-Marom2012}.
The wave function of Bohmian mechanics is a real field permeating the universe, that obeys the standard Schrödinger equation, or more precisely its special-relativistic generalization. 
This latter generalization enables spin properties of the system.
In addition to this the described systems of Bohmian mechanics, often denoted particles, have definite positions and follow continuous trajectories, that gives a time evolution according to the so-called guiding equation
\begin{equation}
     \mathbf v_k=\frac{d\mathbf r_k}{dt}=\mathrm{Im}\frac{\hbar}{m}\left.
     \frac{\langle\psi|\nabla_k|\psi\rangle}
     {\langle\psi|\psi\rangle}\right|_{\mathbf r_i=\mathbf r_i(t)\forall i\ .}
    \label{eq:guiding_equation}
\end{equation}
Note that $|\psi\rangle$ contains spin degrees of freedom, while the particle positions $\mathbf r_k$ do not.
This equation implies that a Bohmian particle in some moving wave packet will travel with the same velocity as the group velocity of the wave packet \cite{Naaman-Marom2012}. 
A simplification occurs when the wave function is a superposition of two orthogonal wave packets $|\psi\rangle=|\psi_A\rangle+|\psi_B\rangle$ that have associated well defined velocities $\mathbf v_A$ and $\mathbf v_B$, then for a single-particle system
\begin{equation}
     \mathbf v=\frac{d\mathbf r}{dt}
     =\frac{\langle\psi_A|\psi_A\rangle\mathbf v_A
     +\langle\psi_B|\psi_B\rangle\mathbf v_B}
     {\langle\psi_A|\psi_A\rangle+\langle\psi_B|\psi_B\rangle},
    \label{eq:guiding_equation_simplified}
\end{equation}
so the resulting velocity is the weighted average of the two constituent velocities \cite{Naaman-Marom2012}. 
This will simplify our calculations below.

Both the Schrödinger equation and the guiding equation describe deterministic evolution. The stochastic outcomes of QM experiments are accounted for in Bohmian mechanics by a lack of knowledge of the initial state of the system. The quantum equilibrium hypothesis (QEH) prescribes that the particle positions at $t=0$ are distributed according to
\begin{equation}
    P\bigl(\mathbf r_1(t)=\mathbf r_1, \ldots,\mathbf r_N(t)=\mathbf r_N\bigr)
    ={\langle\psi|\psi\rangle}\big|_{\mathbf r_i=\mathbf r_i(t)\forall i\ .}
    \label{eq:QEH}
\end{equation}
This is often taken to be a postulate of Bohmian mechanics (although there are exceptions \cite{Durr2013, Durr1992}).
It is by virtue of the QEH that the otherwise deterministic formulation of Bohmian mechanics can produce the same measurement outcomes as standard QM. 
The consequence is a lack of knowledge of the initial position, and then it follows from the Schrödinger and guiding equations that the QM predictions are reproduced by Bohmian mechanics. Nothing further has to be added to the theory, there is no need for further postulates like the collapse of the wave function for when a measurement is performed \cite{Norsen2014}. 

The most straightforward type of measurement in Bohmian mechanics is measuring particle position.
As such, the measurement does not just reveal an already existing position of the particle, but the measurement in a way makes the particle have the detected position. 
Although particles have definite positions in Bohmian mechanics before measurement, a measurement in Bohmian mechanics do not just reveal those positions of the particles. 
Instead, the measurement process will interact with the particles and the wave function, bringing about new positions.

Similarly, the wave function of Bohmian mechanics will not collapse as in ordinary quantum mechanics, but will still change due to the interaction between particle and measuring apparatus. 
There is no way to passively measure a quantum system according to Bohmian mechanics. 
Instead, measurement necessarily requires interaction between the system being measured and the measurement apparatus. 
As such, descriptions of measurements in Bohmian mechanics cannot only involve the wave function and particles of the subsystem being measured but must also include the  measuring apparatus \cite{Durr2013, Durr1992}.

\section{Spin measurement in Bohmian mechanics}
\label{sec:spin}

In this paper we want to study measurement of spin, and in particular restricted to spin-$\frac12$ systems, a restriction which we will make for the remainder of this paper.
For this purpose we need a device that converts spin to position: a Stern-Gerlach (SG) device. 
The presentation here will follow closely that of Refs.~\cite{Norsen2014,Naaman-Marom2012}.
An SG device is a magnet arrangement that will give a traveling particle with spin a change in momentum transverse to the propagation direction, so that different spin components separate into separate paths in the plane spanned by the original propagation direction and the main axis of the magnet arrangement.
In what follows, the propagation direction before the SG device is denoted $x$ and the magnet arrangement axis is denoted~$z$. 

\begin{figure}
    \centering
    a)\includegraphics[scale=.6]{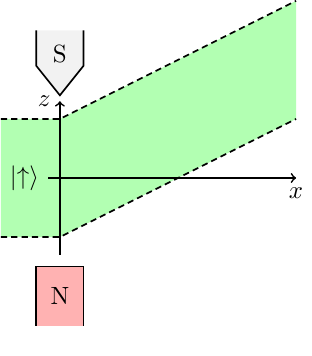}
    \quad
    b)\includegraphics[scale=.6]{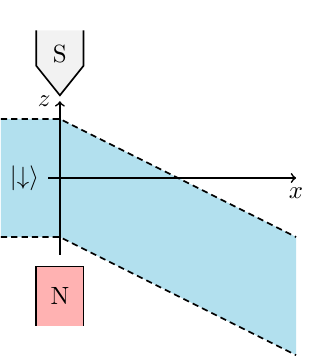}
        
    c)\includegraphics[scale=.6]{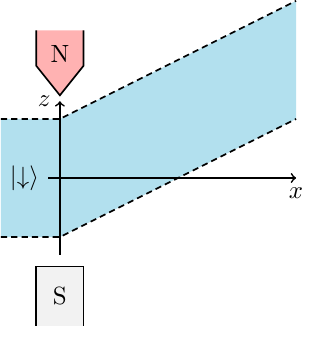}
    \quad
    d)\includegraphics[scale=.6]{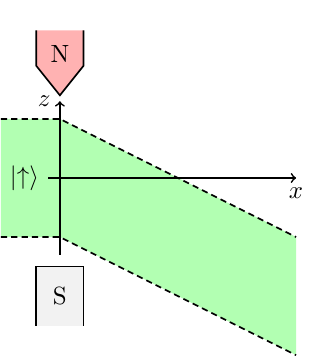}
    \caption{Spin system behavior at SG device, with spatial wave-function according to the main text. 
    a) Spin up system at an SG$_S$ device.
    b) Spin down system at an SG$_S$ device.
    c) Spin down system at an SG$_N$ device.
    d) Spin up system at an SG$_N$ device.
    }
    \label{fig:SG}
\end{figure}

The magnet arrangment could in principle have either a south pole or a north pole in the upper magnet.
In this paper, we will use the value assignment \textit{spin up} for particles that exit the upper path from a SG device with a south pole in the upper magnet, such a device will be denoted SG$_S$.
If we can predict in advance that the particle will exit from the upper path with probability 1, we assign the state $\ket\uparrow$ to the quantum system, see Fig.~\ref{fig:SG}a, and likewise for predicted exit in the lower path assign the state $\ket\downarrow$, see Fig.~\ref{fig:SG}b.
An SG device with a reverse polarity, with a north pole in the upper magnet and denoted SG$_N$,  will instead make particles in the $\ket{\uparrow}$ state exit in the lower output path, see Fig.~\ref{fig:SG}c, and particles in the $\ket{\downarrow}$ state exit in the upper output path, see Fig.~\ref{fig:SG}d. 

If the state is the superposition 
\begin{equation}
    \ket\psi=\alpha\ket{\uparrow}+\beta\ket{\downarrow}
    \label{eq:superposed}
\end{equation}
and an SG$_S$ device is used, the particle will exit the upper path with probability $|\alpha|^2$ and the lower path with probability $|\beta|^2$, normalized so that the total probability $|\alpha|^2+|\beta|^2=1$. 
For an SG$_N$ device, the probabilities are interchanged.

To perform a more Bohmian analysis of an SG device we will use the simplifying assumptions of Ref.~\cite{Naaman-Marom2012}, that the spatial part is constant on a rectangle
\begin{equation}
    \chi(x,z) \equiv \begin{cases}
      \frac{1}{\sqrt{\epsilon a}} , & |x| < \frac{\epsilon}{2}, |z|< \frac{a}{2}\\
      0, & \text{otherwise}, 
    \end{cases}
    \label{eq:spatial_wave_packet}
\end{equation}
with $\epsilon$ small, and that the dynamics has no spatial dependence in the $y$ direction so this can be left out. 
The wave packet has velocity $v$ in the $x$-direction, reaches the center of an SG$_S$ device at time $t=0$, and has the spin component superposition $\alpha \ket{\uparrow} + \beta \ket{\downarrow}$ so that the total wave packet is
\begin{equation}
    \psi(t)=e^{ik(x-\frac{v}{2}t)}\chi(x-vt,z)(\alpha \ket{ \uparrow } + \beta \ket{ \downarrow }).
    \label{eq:initial_wave_packet}
\end{equation}

Since a Bohmian particle in a wave packet will travel with the same velocity as the wave packet, it will have position $r(t)=(vt,z_0)$ for $t\le0$. 
At $t=0$ a momentum kick will be applied to the different components of the wave packet along the $z$ direction. 

\begin{figure}
    \centering
    a)\hspace{-4mm}\includegraphics[scale=0.6]{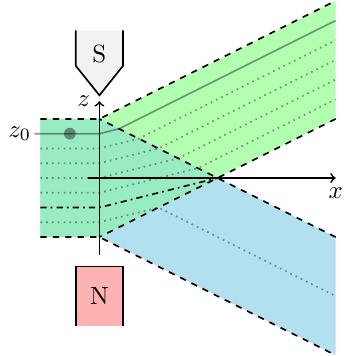}
    \quad
    b)\hspace{-4mm}\includegraphics[scale=0.6]{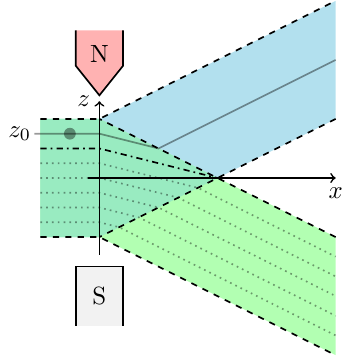}
    \caption{One possible Bohmian trajectory among others for a particle in the state $\ket\psi=\frac{\sqrt3}2\ket\uparrow+\frac12\ket\downarrow$ traveling through an SG device, dash-dotted line is threshold between trajectories going up or down. a)~An SG$_S$ device. b)~An SG$_N$ device.}
    \label{fig:SG_Bohm}
\end{figure}

For an SG$_S$ device, we will take this to result in a transverse velocity of magnitude $+u$ for the $\ket{\uparrow}$ component of the wave packet and $-u$ for the $\ket{\downarrow}$ component \cite{Naaman-Marom2012}. 
The wave function for $t>0$ then will be
\begin{equation}
\begin{split}
    \psi(t)&=\overbrace{e^{ik_{x}(x-\frac{v}{2}t)}e^{ik_{z}(z-\frac{u}{2}t)}\chi (x-vt, z-ut) \alpha\ket{ \uparrow }}^{\psi_A}\\
    & +\underbrace{e^{ik_{x}(x-\frac{v}{2}t)} e^{-ik_{x}(x+\frac{u}{2}t)}\chi (x-vt, z+ut)\beta \ket{ \downarrow }}_{\psi_B}.
\end{split}
    \label{eq:final_wave_packet}
\end{equation}
When the two wave components overlap, the simplified guide Eqn.~(\ref{eq:guiding_equation_simplified}) gives the transverse particle velocity
\begin{equation}
    v_S=\bigl(|\alpha|^2-|\beta|^2\bigr)u.
    \label{eq:velocity_S}
\end{equation}
After this, the wave packets will separate, and the particle will follow one of the wave packets with the same velocity $\pm u$ as the packet in the z-direction, see Fig.~\ref{fig:SG_Bohm}a. 

For an SG$_N$ device, the transverse velocity will be $-u$ for the $\ket{\uparrow}$ component of the wave packet and $+u$ for the $\ket{\downarrow}$ component. 
That is, the $\ket{\uparrow}$-part of the wave packet will now go down, and the $\ket{\downarrow}$-part of the wave packet will now go up after the interaction with the SG$_N$ device. 
Similar calculations as above give the transverse particle velocity in the overlap region of
\begin{equation}
    v_N=\bigl(|\beta|^2-|\alpha|^2\bigr)u=-v_S.
    \label{eq:velocity_N}
\end{equation}
In Fig.~\ref{fig:SG_Bohm} Bohmian trajectories are drawn for several starting positions for $|\alpha|^2=\frac34$, $|\beta|^2=\frac14$ so that $v_S=\frac12u=-v_N$.
The trajectories split at the tip of the overlap region into spin up and spin down trajectories.
For example, for an SG$_S$ device the condition for spin up is
\begin{equation}
z_0>z_S=-av_S/u.
\label{eq:threshold}
\end{equation}
Thus, in Bohmian mechanics wave packets separate and trajectories split as indicated in the example of Fig.~\ref{fig:SG_Bohm}.
One of the separate output paths will have a particle traveling along with it and the other will not, while both wave packets will remain but now be separated in space.

Finalizing the measurement would require measurement of position of the particle, but a full Bohmian treatment of the complete position measurement process would need more consideration.
We will avoid this treatment here and instead use a simpler procedure: blocking one path and proceeding to use the remaining conditional quantum state \cite{Naaman-Marom2012}, under the condition that the particle trajectory is within this conditional quantum state.

\section{Spin is not a property of the Bohmian state alone, if the quantum state is a superposition}

We now look closer at the question if spin is a property of the Bohmian state.
The initial quantum state is identical in Fig.~\ref{fig:SG_Bohm}a and Fig.~\ref{fig:SG_Bohm}b, so the example Bohmian trajectory with initial transverse coordinate $z_0$ will have identical initial Bohmian state in Fig.~\ref{fig:SG_Bohm}a and Fig.~\ref{fig:SG_Bohm}b.
For the superposed example $\ket\psi$ in Fig.~\ref{fig:SG_Bohm}a, if the value $z_0$ is larger than~$z_S=-\frac12a$, the particle will be directed upward.
For the same $\ket\psi$ in Fig.~\ref{fig:SG_Bohm}b, if the value $z_0$ is larger than~$z_N=\frac12a$, the particle will be directed upward.
So in both examples, if $z_0\ge\max(z_S,z_N)=\frac12a$ the trajectory will exit the SG device upward.
Indeed, when a device within Bohmian mechanics splits a family of trajectories, the trajectories directed upward must have originated in the upper part of the family. 
The underlying reason is one of the more prominent features of Bohmian mechanics, that trajectories cannot cross \cite{Bohm1952,Holland1993,Wyatt2005,Teufel2009}.

\begin{figure}
    \centering
    a)\hspace{-4mm}\includegraphics[scale=0.6]{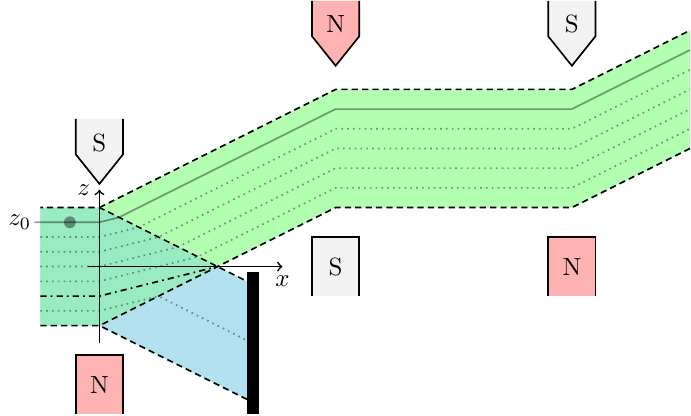}

    b)\hspace{-4mm}\includegraphics[scale=0.6]{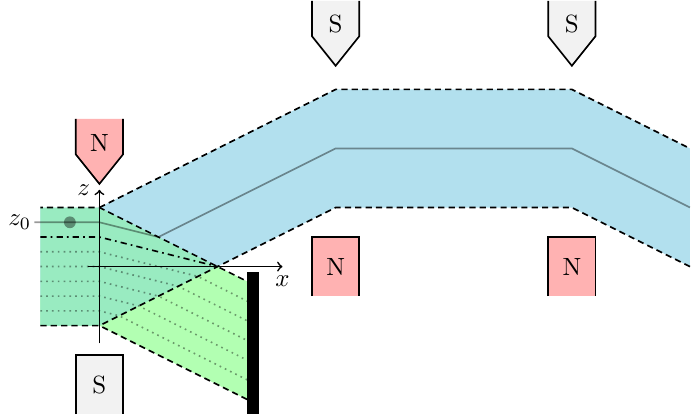}
    \caption{One possible Bohmian trajectory among others for a particle in the state $\frac12\ket\uparrow+\frac{\sqrt3}2\ket\downarrow$ traveling through three subsequent SG devices, with identical SG$_S$ devices as the final measurement. 
    The second SG device is identical to the first except for the inverted polarity, which gives the opposite momentum kick to the wave packet. 
    The lower path out from the first device is blocked and can be left out of consideration.
    a) The last SG$_S$ device verifies that the first SG$_S$ device redirects only "spin up" particles to the upper path.
    b) The last SG$_S$ device verifies that the first SG$_N$ device redirects only "spin down" particles to the upper path.
    This tells us that the Bohmian position $z_0$ together with the wave function preceding the SG device does not determine the outcome of a spin measurement on the system.}
    \label{fig:SG_repeated}
\end{figure}

But the upward wave packet is not the same in Fig.~\ref{fig:SG_Bohm}a and b.
In one case the wave packet has spin component $\ket\uparrow$, and in the other case it has spin component $\ket\downarrow$.
The question is, after blocking the downward wave packet, can we interpret the resulting conditional state as a state with definite spin, and in that case which spin?
This would seem to be simple since the conditional wave packet has a well-defined single spin component, but for all we know, a full Bohmian treatment may make the behavior more complicated.
Checking how the full Bohmian conditional output state behaves is best done by another spin measurement, this time using the same SG$_S$ measurement, identical in both cases.

This is depicted in Fig.~\ref{fig:SG_repeated}: the first SG device is either an SG$_S$ device (Fig.~\ref{fig:SG_repeated}a) or an SG$_N$ device (Fig.~\ref{fig:SG_repeated}b).
The second SG device is identical to the first in both cases, but with reverse polarity to invert the momentum kick of the first SG device, so that the conditional wave packet and particle trajectory are horizontal again.
The third SG device is an SG$_S$ device, indentical in both cases Fig.~\ref{fig:SG_repeated}a and \ref{fig:SG_repeated}b.
If the first SG device is an SG$_S$ device, the third device redirects the wave packet and particle trajectory in the same way as the first. 
Therefore, the conditional Bohmian state output from the first SG$_S$ device of Fig.~\ref{fig:SG_repeated}a indeed can be interpreted as having spin up.
Similarly, the conditional Bohmian state output from the first SG$_N$ device of Fig.~\ref{fig:SG_repeated}b can indeed be interpreted as having spin down.

A Bohmian particle with the same starting position and the same wave packet and wave function in Bohmian mechanics will be assigned a different spin depending on the choice of measurement apparatus used to measure its spin. 
The measurement then does not just ``read off'' some spin value that the particle has before the measurement. 
This argument uses SG devices rather than beamsplitters as used in Ref.~\cite{Albert1992} but is otherwise similar; the present argument strengthens earlier results \cite{Naaman-Marom2012} by using an explicit but unchanging SG measurement to check the spin of the system in the first measurement.

Note that in both cases, the measurement outcome is well defined and can be predicted with probability one: the particle will exit in the upper output beam of the first device if and only if the particle trajectory is above the threshold defined by the wave packet.
But the two choices of SG devices will output different spin particles in the upper beam.
The spin value that will be assigned to the particle depends both on its starting position and wave function, i.e., the full Bohmian state \textit{and} the choice of the measurement apparatus. 
Because of this dependence of the measurement device for the assignment of spin, we cannot assign a definite spin value to the system before it has been measured in the first SG device.
Assigning either spin value to the initial state of the system would either contradict the outcome of the first SG$_S$ device or the first SG$_N$ device. 
We can conclude that the spin value of the particle is not contained in the Bohmian state alone, confirming our assertion in the section title.

\section{The Contextual Ontological Model (COM)}

There is a subtheory of spin-system QM called \textit{stabilizer quantum mechanics} \cite{Gottesman1998b,Aaronson2004a} in which states and measurements are written down in a different notation.
Instead of writing down a specific spin state vector, e.g., $\ket{\uparrow}$ we could use the observable 
\begin{equation}
Z=\left|\uparrow\rangle\langle\uparrow\right|-\left|\downarrow\rangle\langle\downarrow\right|
\end{equation}
associated with spin measurement in the $z$ direction, and describe our state as the one in the $+1$ eigenspace of the observable.
The standard term that is used to express this is that $\ket{ \uparrow } $ is \textit{stabilized by $Z$}.
Similarly, the $\ket{\downarrow}$ state is stabilized by $-Z$, and the positive-phase equal superposition of the two, $\ket\rightarrow=\frac1{\sqrt2}\bigl(\ket\uparrow+\ket\downarrow\bigr)$, meaning $\alpha=\beta$ in the superposed state~(\ref{eq:superposed}) is stabilized by the observable
\begin{equation}
X=\left|\rightarrow\rangle\langle\rightarrow\right|-\left|\leftarrow\rangle\langle\leftarrow\right|,
\end{equation}
associated with a spin measurement in the $x$ direction. 
Finally the negative-phase equal superposition $\ket\leftarrow=\frac1{\sqrt2}\bigl(\ket\uparrow-\ket\downarrow\bigr)$ is stabilized by $-X$.
Extending this to several spin systems will give the full subtheory where, e.g., $\ket{\uparrow}\otimes\ket\uparrow$ is stabilized by $Z\otimes\mathbb{I}$ and $\mathbb{I}\otimes Z$.
The subtheory does not contain all quantum states, but does contain some entangled states such as the Bell singlet state $\frac1{\sqrt2}\bigl(\ket\uparrow\otimes\ket\uparrow -\ket\downarrow\otimes\ket\downarrow\bigr)$ \cite{Bell1964}, the unique state stabilized by both $-Z\otimes Z$ and $-X\otimes X$.
This makes the subtheory highly nontrivial and interesting to study \cite{Gottesman1998b,Aaronson2004a}.

Writing the state in terms of stabilizers is equivalent to writing down the predicted measurement outcomes for the associated observables.
If the measured state is stabilized by $Z$, the measurement outcome when measuring the $Z$ observable is $+1$, while if the measured state is stabilized by $-Z$, the measurement outcome when measuring $Z$ is $-1$. 
If instead the measured state is stabilized by an observable that does not commute with the measurement, e.g., $X$, the outcome is random equally probable, mirroring the predictions from the vector description $\frac1{\sqrt2}\bigl(\ket\uparrow+\ket\downarrow\bigr)$. 
The state is then updated to an appropriate stabilizer, here $+Z$ if the outcome was $+1$ or $-Z$ if the outcome was $-1$, also mirroring the quantum state update to $\ket\uparrow$ or $\ket\downarrow$.
This all becomes much more complicated and interesting in the many-system case, but in this paper we will concentrate on repeated measurement on a single system.

Stabilizer QM can now be extended so that outcomes of both $Z$ and $X$ measurements can be predicted with probability given the extended state of the system.
In COM \cite{Hindlycke2022,Larsson2026} an extension of stabilizer quantum mechanics is created that assigns definite outcomes not only to the stabilizers of the state but also to observables that do not commute with the stabilizers, whose outcomes were random in stabilizer quantum mechanics.
For this purpose, the model uses \textit{destabilizers}, observables that do not commute with the stabilizers and also do not stabilize the state, but still encode measurement outcomes determined in addition to the outcomes specified by the QM state.
For a single spin the model is reasonably simple. 
The COM states that correspond to $\ket\uparrow$ are
\begin{equation}
\{Z;X\}\text{\quad or \quad}\{Z;-X\}.
\end{equation}
with equal probability. The first element is the stabilizer, the second is the destabilizer, and the COM state fixes the outcomes both for a $Z$ and an $X$ measurement.

Our spin measurement in the previous sections corresponds to a $Z$ measurement on a quantum state stabilized by $X$, if $\alpha=\beta$.
Again COM would use either the destabilizer $+Z$ or $-Z$ with equal probability, so the COM state would either be
\begin{equation}
\{X;Z\}\text{\quad or \quad}\{X;-Z\}.
\end{equation}
Comparing with Bohmian mechanics, the quantum spin state is specified by the stabilizer $X$, while the sign $z_0=\pm1$ of the $Z$ operator corresponds to the Bohmian position coordinate and controls the outcome.

Measurement of $Z$ on this COM state now consists of three steps. 
The first is to report the measurement outcome $z_0=+1$ or $z_0=-1$ depending on the destabilizer sign $z_0Z=\pm Z$, the second is to make the former destabilizer $z_0Z$ become the stabilizer of the new state, and third is to make the former stabilizer become the destabilizer of the new state and randomize the sign, creating the new state 
\begin{equation}
\{z_0Z;X\}\text{\quad or \quad}\{z_0Z;-X\}.
\end{equation}
In COM, a repeated measurement of $Z$ will output $z_0$ again, but since we are now reading off the stabilizer in step 1, no replacement will happen in step 2, and the state update for this repeated measurement in step 3 will only re-randomize the sign of the destabilizer $X$.
The many-system state update is much more complicated \cite{Hindlycke2022,Larsson2026} but this suffices for our needs in the present paper.

\section{Spin is a property of the COM state alone, even when the quantum state is a superposition}

In our SG devices we associate the $+1$ measurement outcome with the upper output path, and the $-1$ measurement outcome with the lower output path. 
This is consistent with measurement on the $\ket\uparrow$ and $\ket\downarrow$ states, stabilized by $\pm Z$, because of the following.
An SG$_S$ device performs a $Z$ measurement, which connects the ``spin up'' outcome with the upper output path labeled $+1$, and the ``spin down'' outcome with the lower output path labeled $-1$, this reproduces the behavior in Fig.~\ref{fig:SG}a-b.
Similarly, an SG$_N$ device performs a $-Z$ measurement, which instead connects the  ``spin up'' outcome with the lower output path labeled $-1$, and ``spin down'' outcome with the upper output path labeled $+1$, this reproduces the behavior in Fig.~\ref{fig:SG}c-d.

The superposed state that we use in Fig.~\ref{fig:SG_Bohm} (with $\alpha=\beta$) has the COM state
\begin{equation}
\{X,z_0Z\}
\end{equation}
where $z_0=\pm1$ is random equally distributed.
Since an SG$_S$ device performs a $Z$ measurement, it will here output $z_0$ reproducing the behavior in Fig.~\ref{fig:SG_Bohm}a, and update the state to
\begin{equation}
\{z_0Z,\pm X\}.
\label{eq:Z-post-measurement}
\end{equation}
On the other hand, an SG$_N$ device will perform a $-Z$ measurement, which will output $-z_0$ because the phase of the observable $-Z$ is obtained from the destabilizer as $-z_0(-Z)=z_0Z$, this reproduces the behavior in Fig.~\ref{fig:SG_Bohm}b. 
The state will update to
\begin{equation}
\{-z_0(-Z),\pm X\}=\{z_0Z,\pm X\}.
\label{eq:-Z-post-measurement}
\end{equation}
Note that the post-measurement stabilizer is the same in Eqn.~(\ref{eq:Z-post-measurement}) and Eqn.~(\ref{eq:-Z-post-measurement}).
Therefore, in both cases, a second $Z$ measurement will give the outcome $z_0$.
The outcome of the second $Z$ measurement, in both cases, only depends on the initial value of $z_0$, it does not depend on if the intermediate measurement uses an SG$_S$ device or an SG$_N$ device.
This behavior is completely different from the Bohmian behavior in Fig.~\ref{fig:SG_repeated}a-b where the outcome of the second measurement depends both on the initial value of $z_0$ and on the choice of intermediate measurement.
We can conclude that the spin value of the particle is contained in the COM state alone, confirming our assertion in the section title.

\section{Conclusion}

In this paper, we have compared the kind of contextuality occuring in Bohmian mechanics \cite{Albert1992,Hardy1996,Norsen2014} with Kochen-Specker (KS) contextuality \cite{Kochen1967,Budroni2022} and shown that they are not the same, and in particular shown that Bohmian contextuality is not necessary for KS contextuality despite claims to the contrary. 

In Bohmian mechanics measurement outcomes cannot always be predicted with certainty from the system state alone \cite{Albert1992,Hardy1996,Norsen2014}. 
The presentation here uses sequential Stern-Gerlach devices to show that spin measurement outcomes cannot be predicted from the Bohmian state only, indeed that spin measurement outcomes can only be predicted from knowledge of the joint system of the measurement device and the Bohmian state of the wave packet and particle. 
Spin can therefore not be considered an intrinsic property of the Bohmian state, but emerges in the interaction between the system and the measurement apparatus, as has been previously reported.

We then showed that the Contextual Ontological Model (COM) \cite{} gives a fundamentally different picture than that of Bohmian mechanics. 
In this latter model, measurement outcomes of spin are determined directly by the state itself, without any dependence on the measurement setup. 
Within the model, the values of the observables can be considered to exist before the measurement and are as such simply revealed during measurement. 
Bohmian contextuality of the type described above is therefore not present in COM but the model still reproduces KS contextuality. 
Therefore the form of contextuality found in Bohmian mechanics, where measurement outcomes depend on the measurement configuration, is not required for KS contextuality in every hidden variable model that reproduces the predictions of quantum mechanics. 
There are several such notions that are easy to conflate \cite{Budroni2022}, but our result here shows that at a minimum the two notions of contextuality treated in this paper are separate.

This is interesting in itself but also tells us that COM is not just a simplified version of Bohmian mechanics, it is an independent hidden-variable model that could be developed further as an alternative ontological version of quantum mechanics. 
The main conclusion in this paper is that even though quantum mechanics does require KS contextuality in hidden variable models, it does not require the precise form of contextuality that occurs in Bohmian mechanics.

\begin{acknowledgments}
This work is supported by the Swedish Research Council project no 2023-05031.
\end{acknowledgments}

\bibliographystyle{apsrev4-2}
\bibliography{bohmian-mechanics-is-not-com}

\end{document}